\documentclass[aps,prd,twocolumn,showpacs,floatfix]{revtex4}
\newcommand{\be}{\begin{equation}}
\newcommand{\ee}{\end{equation}}
\newcommand{\jcap}{J. Cosmol. Astrophys.} 
\newcommand{\mnras}{Mon. Not. Roy. Astr. Soc.}
\newcommand{\npb}{Nucl. Phys. B}
\begin{document}

\title{Are two kinds of dark matter seen in Galactic gamma rays?}

\author{Matts Roos} 
\affiliation{Department of Physical Sciences, Division of High Energy Physics,
University of Helsinki, Helsinki}

\date{\today}

\begin{abstract}

Excesses of Galactic gamma rays in the 1-100 GeV region observed by EGRET and by WMAP have been interpreted as signals of dark matter (LSP) annihilation. We argue that the excess of TeV gamma rays from the Galactic center observed by H.E.S.S. signals a heavier dark matter component with a very small relic density.

\end{abstract}
\pacs{95.35.+d, 14.80.Ly, 98.35.Jk, 98.70.Vc}
\maketitle

\section{Introduction}
Dark matter has so far manifested itself only through its gravitational effects -- in fact on several scales in different ways. If dark matter is composed of stable, massive neutral particles having no or very weak interactions with baryonic and leptonic matter, WIMPs, it may be very difficult to detect them unless they annihilate into photons, either directly or via quark-antiquark pairs or lepton-antilepton pairs. The popular Supersymmetric (SUSY) models \cite{SUSY} offer several candidate particles, in particular the lightest SUSY particle (LSP) which must be stable.

In our Galaxy dark matter is abundant, forming of a halo \cite{Klypin} with some radial density distribution. Until particle physics experiments clarify the nature of dark matter, the best source of information may be annihilation gamma rays from the Galaxy center (GC), from the disk, the spiral arms, and from the halo. Indeed, several recent observations of gamma rays in different energy ranges \cite{EGRET,WMAP,HESS} have been analyzed and interpreted as signatures of LSP dark matter self-annihilation \cite{Cesarini,Finkbeiner,de Boer,Horns,Silk}. 

\section{EGRET}
The diffuse Galactic $\gamma$-ray spectrum observed by the EGRET satellite extends from about 30 MeV to almost 100 GeV \cite{EGRET,Strong}. Below 1 GeV the spectrum is fairly well described  by conventional sources: energetic baryonic interactions in the ISM resulting in secondary $\pi^0$ decays, inverse Compton (IC) scattering of electrons on starlight and on CMB photons, and Bremsstrahlung from electrons 
\cite{Strong,Cesarini,de Boer}. Above 1 GeV the EGRET spectrum shows a significant excess over the conventional sources, with the same spectral shape in 6 sky directions, as is expected for dark matter annihilation \cite{de Boer} throughout the Galaxy. Various explanations modifying one conventional component but not invoking dark matter have been tried (referenced and discussed in \cite{Strong}), but either the baryonic component has to originate from a harder proton spectrum, in conflict with local limits on antiproton and positron abundances, or the Compton scattering electrons have to be harder than what is locally observed, or one has to adjust the absolute normalizations of all the three components {\em ad hoc}.

Invoking the presence of dark matter, a hard electron spectrum could be obtained from LSP annihilations $\chi\bar{\chi}$, mainly into $b \bar{b}$ quark pairs. Synchrotron radiation from such ultrarelativistic electrons, as evaluated with the help of the DarkSusy program \cite{DarkSusy} and the Pythia program \cite{Pythia}, turns out to reproduce the shape of the EGRET excess spectrum well for LSP masses in the range $m_\chi=50-90$ GeV \cite{de Boer}.

However, to obtain the absolute normalization in each of the 6 directions one would need a model for the triaxial halo profile of dark matter densities. Although there exist phenomenological profiles derived by averaging over hundreds of galaxies \cite{NFW,Moore,Jimenez}, the average profile is not necessarily a good description of our Galaxy. Inversely, one can use the observed excess spectrum to trace the profile: a spherical WIMP halo radiates isotropically, the higher star density in the disk testifies to an enhanced WIMP density there, and the GC density may be more or less cuspy. De Boer et al. \cite{de Boer} propose a disk model boosted with two radial rings of enhanced density, the inner one fitted at radius 4.3 kpc and the outer one at $14\pm 2.1$ kpc. It is intriguing to note that this model is supported by the observation of visible matter forming a ring of stars at $14-18$ kpc with enhanced density. This two ring model also fits the rotation curves well, including the mysterious change of sign of the slope at 9.4 kpc (for references see \cite{de Boer}).

Thus the EGRET excess appears to be well explained by $\gamma$-rays from the self-annihilation of LSPs \cite{de Boer} with a relic density in consistency with the first year result of WMAP \cite{Spergel}, $\Omega_\chi = 0.23\pm 0.04$, and with a total annihilation cross-section of $\langle\sigma_A v\rangle\sim 2\cdot 10^{-26}$ cm$^3$/s .

\section{WMAP}
Let us now turn to another window. In a careful analysis of Galactic foreground components in the WMAP data, Finkbeiner \cite{Finkbeiner} has noted an excess emission from the inner Galaxy, distributed with radial symmetry within $20^\circ -30^\circ$ around the GC. He argues that this may represent inverse Compton scattered starlight on diffusion hardened electrons, produced by annihilation of $\sim 100$ GeV WIMPs into $e^+ e^-$ pairs. 

To analyze this radiation Finkbeiner \cite{Finkbeiner} assumed a smooth spherical density distribution with a truncated NFW profile \cite{NFW} in the inner 1 - 2 kpc Galaxy, and evolved the electron spectrum through diffusion and IC energy loss. 

The resulting $\gamma$-ray spectrum appears to have a shape rather similar to that of the EGRET excess (an exact comparison is rendered difficult by the different units used in \cite{Finkbeiner} and \cite{de Boer}), thus supporting the interpretation of LSP annihilation. The two analyses differ clearly in their density profiles in the galactic plane, and also in the annihilation mechanism: the power is assumed either to go primarily into $b \bar{b}$ \cite{de Boer} or entirely into ultrarelativistic $e^+ e^-$ pairs \cite{Finkbeiner}. One can still conclude that the two studies observe the annihilation of the same particle $\chi$.

\section{H.E.S.S.}
Aharonian et al. \cite{HESS} have announced the detection of very high energy (VHE) $\gamma$-rays from a point-like source coincident within 1' of Sagittarius A* in the GC, obtained with the H.E.S.S. array of Cherenkov telescopes. The observed energy spectrum ranges from 0.3 TeV to 10 TeV and is rather flat. 

Although various conventional sources of VHE $\gamma$-rays may still be plausible explanations \cite{HESS}, we shall take the attitude that the source is annihilation of a heavier WIMP, $\xi$, which cannot be the source of the much softer $\chi$ annihilation $\gamma$-rays seen by EGRET and WMAP. If $\chi$ is the LSP, as has been argued above and in \cite{de Boer,Finkbeiner}, the  much heavier $\xi$ would have frozen out of thermal equilibrium earlier during the radiation-dominated epoch. 

Let us take the $\xi$ mass to be $m_\xi$, the relic density parameter $\Omega_\xi$ and the average annihilation cross-section $\langle\sigma_A v\rangle$. The $\xi$ fall out of thermal equilibrium at temperature $T_{eq}\ll m_\xi$, and they annihilate until their number density $n_\xi$  has been reduced to give $\langle\sigma_A v\rangle n_\xi \approx H$. The properties of the $\xi$ are constrained by the relations 
\be
\Omega_\xi\propto\langle\sigma_A v\rangle^{-1}\ \ \ \ \ 
\langle\sigma_A v\rangle\propto m_\xi^{-1}\ .
\ee
\noindent Using the WMAP value for the $\chi$ relic density $\Omega_\chi$, a value for $m_\chi$ can be derived. Now, however, there is not room for a high $\xi$ density as well as the high $\chi$ density, so we must have $\Omega_\xi\ll\Omega_\chi$. In fact, the order of magnitude is set by the error on the $\Omega_\chi$ measurement \cite{Spergel}, thus $\Omega_\xi\leq 0.04$.

This relic density  has already been estimated by Silk \cite{Silk} for a typical NFW profile \cite{NFW} and an inferred cross-section of 10 pb at the upper end of the H.E.S.S. energy spectrum, 10 TeV, with the result $\Omega_\xi\approx 0.03$. This is exactly the value needed, although it has met with great difficulties in previous analyses \cite{HESS,Horns,Silk} which treated the H.E.S.S. detection as LSP annihilation, not recognizing the r\^{o}le of the EGRET and WMAP excesses. 

Two-WIMP models that have lately been discussed are, for instance, by Boehm and Fayet \cite{Boehm}, who consider theories displaying an underlying $N=2$ extended supersymmetry, and by Hooper and March-Russell \cite{Hooper}, who take a stable dark matter candidate from the messenger sector of gauge mediated supersymmetry breaking models. With a mass of 20 to 30 TeV they can indeed generate the spectrum seen by HESS with the WMAP relic density. They probably possess enough freedom also to match a very small relic density, as we have advocated here.

\section{DISCUSSION}
This two-WIMP conjecture has several interesting consequences. Knowing the approximate masses and annihilation cross-sections of two neutral SUSY particles is a considerable new input to SUSY spectroscopy. 

Having two dark matter particles with so different masses changes the picture of galaxy density profiles. The $\chi$ need not have much of a central cusp, whereas the $\xi$ probably dominates the innermost region. This would feedback to the EGRET analysis which needed an inner ring of dark matter \cite{de Boer}, as well as to the WMAP analysis which used a truncated NFW profile \cite{Finkbeiner}.

The picture is considerably broadened if one also considers schemes of coannihilation. Also, some of the charged $\xi$ annihilation products may perhaps diffuse out of the GC and spill lower energy photons into the upper ends of the EGRET and WMAP spectra, thus affecting the analyses at hand. There is clearly need for a simultaneous study of the EGRET excess and the WMAP excess.

\end{document}